# Space Weather Application Using Projected Velocity Asymmetry of Halo CMEs

G. Michalek[1], N. Gopalswamy[2], S. Yashiro[3]

[1] *Astronomical Observatory of Jagiellonian University, Cracow, Poland (michalek@oa.uj.edu.pl)*
[2] *Solar System Exploration Division, NASA GSFC, Greenbelt, Maryland*
[3] *Center for Solar and Space Weather, Catholic University of America*

**Abstract.**
Halo coronal mass ejections (HCMEs) originating from regions close to the center of the Sun are likely to be responsible for severe geomagnetic storms. It is important to predict geo-effectiveness of HCMEs using observations when they are still near the Sun. Unfortunately, coronagraphic observations do not provide true speeds of CMEs due to the projection effects. In the present paper, we present a new technique allowing estimate the space speed and approximate source location using projected speeds measured at different position angles for a given HCME (velocity asymmetry). We apply this technique to HCMEs observed during 2001-2002 and find that the improved speeds are better correlated with the travel times of HCMEs to Earth and with the magnitudes ensuing geomagnetic storms.

## 1. Introduction

Halo coronal mass ejections (HCMEs) originating from regions close to the central meridian of the Sun and directed toward Earth cause the most severe geomagnetic storms (Gopalswamy, Yashiro, and Akiyama, 2007 and references therein). Using coronagraphic observations it is possible to determine occurrence rate of CMEs, direction of propagation in the plane of the sky, angular width, and speed (*e.g.* Kahler, 1992; Webb *et al.*, 2000; St. Cyr *et al.*, 2000, Gopalswamy, Lara, and Yashiro, 2003; Gopalswamy, 2004; Yashiro *et al.*, 2004). It is well known that geoffective CMEs originate mostly within a heliographic longitude ±30° (Gopalswamy *et al.*, 2000, 2001; Webb *et al.*, 2000; Wang, Sheeley, and Andrews, 2002; Zhang *et al.*, 2003) and that the initial speed of the CMEs is correlated with the $D_{ST}$ index (Tsurutani and Gonzalez, 1998; Srivastava and Venkatakrishnan, 2002, Yurchyshyn, Wang, and Abramenko, 2004). These studies were based on the plane of the sky speeds of CMEs without considering the projection effects. Recently, there have been several attempts to correct for the projection effects and determine the real CME parameters (Zhao, Plunkett, and Liu, 2002 (ZPL); Michalek, Gopalswamy, and Yashiro, 2003 (MGY); Xie, Ofman, and Lawrence, 2004 (XOL); Michalek, 2006 (M)). These techniques, which need special measurements in coronagraphic fields of view, were based on the assumptions that CMEs have cone shapes and propagate with constant speeds. Based on the cone model, Michalek *et al.* (2006) demonstrated that fast CMEs originating on the western hemisphere close to the disk center are most likely to cause the most severe geomagnetic storms (Dst $\leq$ -150 nT), although there have been some east-hemisphere CMEs capable of causing such kind of storms (Gopalswamy *et al.*, 2005a; Dal Lago *et al.*, 2006). Recently Moon *et al.* (2005) introduced a new direction parameter in correlation with geo-effectiveness of CMEs. It is defined as the ratio of the shortest to the largest distance of the CME front from the disk center for a given LASCO image. They demonstrated that this parameter is strongly correlated with the geo-effectiveness of CMEs. In this paper we present a new but similar technique to calculate the true speed of HCMEs, based on the differences in the projected speeds measured around the occulting disk. Our direction parameters is determined from several (4-5) LASCO images and should be more accurate in describing the asymmetry of HCMEs. We use the space speeds for the prediction of travel times (TT) of HCMEs to Earth and to determine magnitudes ($D_{ST}$ index) of geomagnetic storms. Results obtained with this simple technique are compared with those from other methods (cone models). The paper is organized as follows: in Section 2, the technique used to determine the improved speeds is presented. In Section 3, the data considered for this study is described. In Section 4, we implement the improved speeds for space weather forecasting. Finally, conclusions are presented in Section 5.

## 2. The technique for the speed estimation

The sky-plane measurement describing the properties of CMEs, especially HCMEs, are subject to the projection effects. However, true speeds in the direction of Earth are needed for estimating the travel times of CMEs to Earth (Gopaswamy *et al.*, 2001). This issue is crucial for space weather forecasting. We propose to get realistic speeds by determining an asymmetry ratio (ASR) in the projected speeds measured at different position angles around the occulting disk. The ASR is defined as the ratio between the maximum and minimum speeds measured at different position angles around the occulting disk for a given halo CME. A two-step procedure is carried out to obtain the ASR. First, using height-time plots measurements at every 15° the projected speeds were determined at 24 position angles (see Figure 1 for an illustration of the method). Next, from these values, we calculate running averages from four successive speeds around the occulting disk. This allowed us to get 24 average speeds around the occulting disk for each event. Thus the ASR is the ratio between the maximum ($<V_{max}>$) and minimum ($<V_{min}>$) average seeds determined in the way described above. There are two reasons to use the above procedure in determining ASRs. First, CMEs sometimes are irregular with narrow and fast structures which can over-estimate the measured projected speeds. Second, CMEs can be faint and single measurements are prone to large errors. Our procedure minimizes these errors. It is important to note that the ASR has a real physical meaning. It is commonly assumed that CMEs have a cone shape which was described in details in a few papers (Michalek, Gopalswamy and Yashiro, 2003, Michalek, 2006, Xie, Ofman and Lawrence, 2004). Thus this parameter obtained directly from coronagraphic observations should better indicate CME propagation directions than could be deduced from the associated flare locations. Such locations of CMEs are very useful to eliminate the projection effects and to estimate the radial speeds and geo-effectiveness of CMEs. Moon *et al.* (2005) demonstrated that a direction parameter obtained from the asymmetry in the 3D shape of CMEs has a relatively good correlation with the geo-effectiveness of CMEs. Many halo CMEs have irregular shapes in the LASCO field of view, so the procedure proposed by Moon *et al.* (2005) is not accurate for such events. To minimize errors we propose to determine the directional parameter in the velocity space form several (4-5) LASCO images. Following their consideration, we defined the ASR using the maximum ratio between these projected speeds for a given event. We assume that the improved speeds of CMEs could be expressed by the formula:

$$V_{imp} = <V_{max}> + <V_{max}>/ASR \qquad (1)$$

where

$$ASR = <V_{max}> / <V_{min}> . \qquad (1)$$

From these equations, we see that $V_{imp}$ could be expressed as a simple sum of $V_{max}$ and $V_{min}$. If a given CME originates at the limb then $V_{min} = 0$ and the radial speed is exactly equal to $V_{max}$. If the location of a given CME moves from the limb to the center of the Sun then $V_{max}$ decreases and $V_{min}$ increases. At the disk center, when HCMEs are symmetric, $V_{min} = 2 <V_{max}>$. It is not a perfect approximation of the radial speed but the only possible because other method do not work for symmetric HCMEs. The projection effects are mostly determined by the source locations and widths of CMEs. Unfortunately, in comparison to the cone models, our technique allows us to estimate only the locations of CMEs.

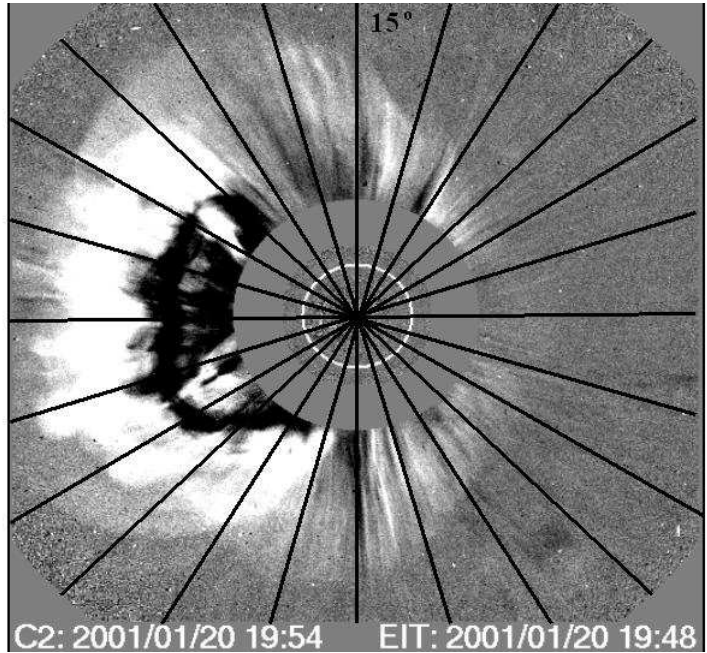

*Figure 1.* An example of HCME (20 Jan. 2001) with dark lines along which the height-time measurements were performed.

## 3. Data

A list of HCMEs studied in this paper is displayed in Table 1. We considered only front-side full HCMEs during a period from the beginning of 2001 until the end of 2002. We selected this limited period of time to get a representative sample of HCMEs to test our new tchnique. In the considered period of time, 70 front-side full HCMEs were found in SOHO/LASCO catalog. One of them was too faint to perform necessary measurements so 69 front-side HCMEs are displayed in the Table 1. Using data from World Data Center Kyoto (http://swdcdb.kugi.kyoto-u.ac.jp) geomagnetic disturbances caused by these events were investigated. In order to find a relationship between HCMEs and magnetic disturbances a two step procedure was used. First, we found all geomagnetic storms, in the considered period of time, with $D_{ST}$ index $\leq -30nT$ (this high limit was chosen following Michalek *et al.* (2006)). We assume that the associated magnetic disturbance should start no later than 120 hours after the first appearance of a given event in LASCO field of view and no sooner than the necessary travel time of a given CME to Earth calculated from the measured maximal projected speed. We related a given disturbance with a given HCME if they were within the specified time range. 20 events from our list were not geoeffective ($D_{ST} > -30 nT$). These HCMEs were slow ($V < 900 km/s$) or originated closer to the solar limb. By examining solar wind plasma data (from Solar Wind Experiment, Ogilvie *et al.*, 1995) and interplanetary magnetic field data (from Magnetic Field Investigation (Wind/MFI) instrument, Lepping *et al.*, 1995), we identified interplanetary shocks driven by the respective interplanetary CMEs (ICMEs). Measuring the time when a HCME first appears in the LASCOs field of view and the arrival time of the corresponding shock at Earth the travel time (TT) can be determined (*e.g.* Manoharan *et al.*, 2004; Gopalswamy *et al.*, 2005c).

The results of our considerations are presented in Table 1. The first column gives the date of the first appearance in the LASCO field of view. Next three columns displays parameters obtained from the projected speeds ($< V_{max} >, < V_{min} >$ and ASR). In column (5) linear speeds from SOHO/LASCO catalog are presented. Location of flares associated with CMEs are shown in column (6). In column (7) the minimum values of $D_{ST}$ indices for geomagnetic disturbances caused by HCMEs are displayed. Finally, in column (8) the travel time (TT) of shock to the Earth are presented. ASR varies from 1.11 to 7.59. The lowest values are for events close to the disk center. The highest values are for limb HCMEs. $V_{max}$ is close to the catalog speed ($V$) for most events. The difference arises from the fact that $V_{max}$ is obtained as an average over 4 PAs, while $V$ is form a single position angle.

## 4. Space weather application

For space weather application it is crucial to predict, with good accuracy, the travel times (TT) and strengths ($D_{ST}$) of geomagnetic disturbances. In the next two subsections, we consider these issues using the improved speeds and the ASR parameter.

### 4.1. Prediction of the onset of geomagnetic disturbances

Figure 2 shows the scatter plot of the plane of the sky speeds (from SOHO/LASCO catalog) versus travel time. Correlation coefficients are 0.68 for the western and 0.50 for the eastern events. The standard error in determination of the TT is 16 hours. For comparison, we present a similar plot except that we use the the improved speeds (Figure 3, left panel). The figures clearly show that the improved speeds are better correlated with the TT, the correlations coefficients more significant: 0.72 for the western and 0.74 for the eastern events. The standard error in TT is 12 hours. The linear correlation coefficients are almost the same as those for the asymmetric cone model (Michalek, Gopalswamy and Yashiro, 2007). For (M) model the standard error was smaller by ∼2 hours. To illustrate that our considerations are consistent with previous results we compare them with the ESA model (the continuous line, Gopalswamy *et al.* 2005b). For these plots we used only the 49 geoeffective ($D_{ST} \leq -30nT$) events. For comparison, in Figure 3 (right panel) we have added the three events (2000/07/14, 2003/10/28, 2003/10/29) of historical importance, represented by the dark diamonds (Gopalswamy *et al.*, 2005b). In our list there are CMEs originating from close to the disk center and all the way to the limb. For arrival at Earth we need to consider Earth-directed speeds rather than the space speeds. We obtained the Earth-directed speeds for the 49 geoeffective events and plotted them in Figure 3 (right panel). We see that points agree well with the ESA model.

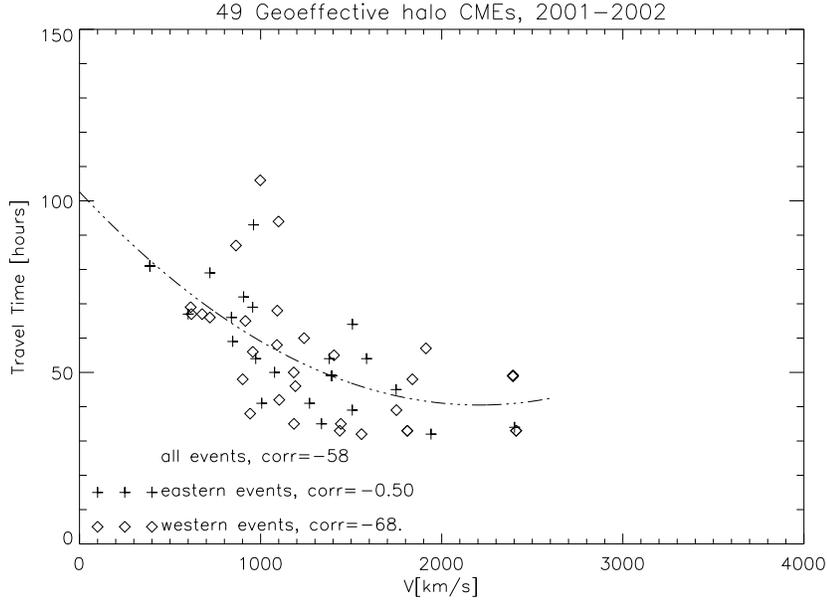

*Figure 2.* Scatter plot of the plane of the sky speeds versus the travel times (TT). Diamond symbols represent events originating from the western hemisphere and cross symbols represent events originating from the eastern hemisphere. The dot-dashed line is a second degree polynomial fit to all the data points.

## 4.2. MAGNITUDES OF GEOMAGNETIC DISTURBANCES

Magnitudes of geomagnetic storms depend not only on the speeds of CMEs but also on the location of source region on the solar disk (*e.g.* Moon *et al.*, 2005; Gopalswamy, Yashiro and Akiyama, 2007). Michalek *et al.* (2006) found a significant correlation between $D_{ST}$ and $V * \gamma$ where the parameter

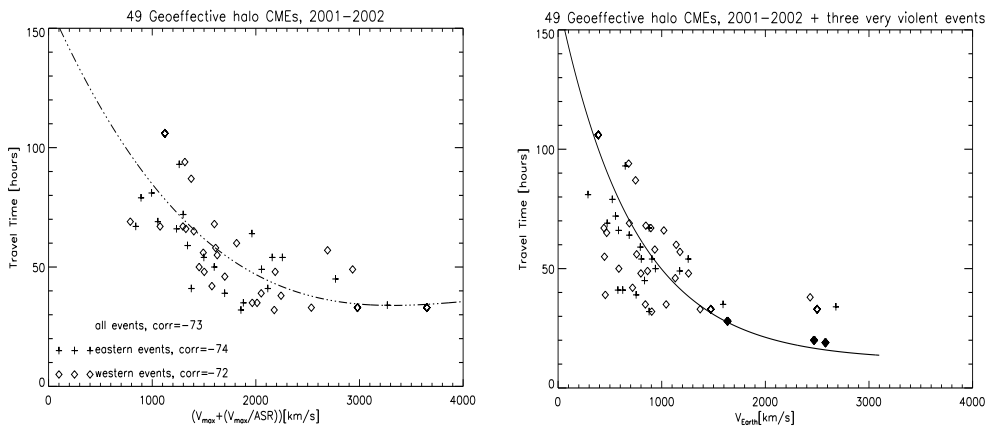

*Figure 3.* The scatter plots of the improved (left panel) and Earth directed speeds (right panel) versus the HCME travel time (TT). Diamond and cross symbols represent events originating from the western and eastern hemispheres, respectively. The dot-dashed line (left panel) is a second order polynomial fit to all the data points. The continuous line (right panel) is the ESA model representation. The three additional dark diamonds (only on the right panel) show the HCMEs (2000/07/14, 2003/10/28 and 2003/10/29) of historical importance.

$\gamma$ is the angular distance of a given CME from the plane of the sky. This parameter decides which part of a HCME hits Earth. Events with small $\gamma$ strike Earth with their flanks while those with large $\gamma$ hit Earth with their central parts. Figure 4 shows the scatter plots of the plane of the sky speeds multiplied by $\gamma$ versus $D_{ST}$ index. The parameter $\gamma$ was determined from location of associated flares. There is a slight correlation between $(V * \gamma)$ and $D_{ST}$. Correlation coefficients are approximately 0.49 for the western and 0.30 for eastern events, respectively. Locations of the associated flares could be different from locations of the associated CMEs. For comparison, Figure 5 shows similar plot but for the improved speeds and the ASR parameters. We propose to use the ASR instead of $\gamma$. This parameter includes source locations of CMEs (not flares) and should be more useful. There is only one difference to $\gamma$. Larger ASR means a location closer to the limb ($\gamma \sim 1/ASR$). Figure 5 shows the scatter plot of $(<V_{max}> + <V_{max}>/ASR)/ASR$ versus $D_{ST}$ index. In other worlds, it is the scatter plot of improved speeds $((<V_{max}> + <V_{max}>/ASR) = V_{imp})$ multiplied by source locations ($1/ASR \sim \gamma$) versus $D_{ST}$ index. We used similar formula in our previous paper (Michalek *et al.*, 2006). There is a significant correlation between $V_{imp}/ASR$ and $D_{ST}$. Correlation coefficients are almost two times larger than those obtained from the projected speeds: -0.65 for the western and -0.61 for eastern events, respectively. It is clear that the our technic could be very useful for space weather applications. For these plots (Figure 4 and Figure 5), we used all HCMEs form Table 1, even the non-geoeffective ones ($D_{ST} > -30 nT$). These events generate false alarms. Non-geoeffective HCMEs are slow ($V < 900$km s$^{-1}$) or have source region closer to the solar limb (see also Gopalswamy, Yashiro and Akiyama, 2007). The correlation coeffcient are slightly less significant in comparison to these received with the asymmetric cone model (Michalek, Gopalswamy and Yashiro, 2007). In the case of the cone model the correlation coefficient for all events was -0.72.

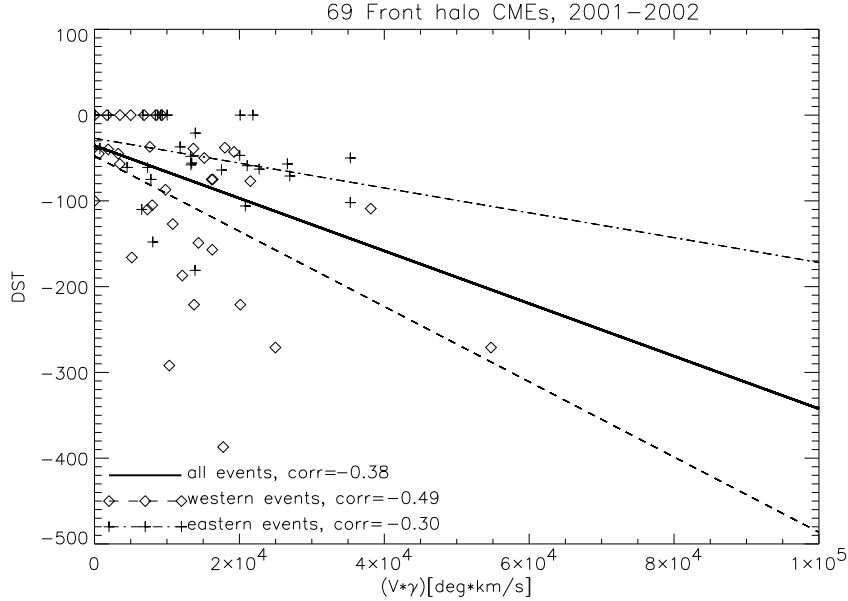

*Figure 4.* The scatter plot of the plane of the sky speeds multiplied by $\gamma$ versus $D_{ST}$ index. Diamond symbols represent events originating from the western hemisphere and cross symbols represent events originating from the eastern hemisphere. The solid line is a linear fit to all the data points, the dot-dashed line is a linear fit to eastern events, and the dashed line is a linear fit western events.

## 5. Summary

HCMEs are likely to cause severe geomagnetic storms. Speed is one of the most important parameter characterizing geo-effectiveness of CMEs. Un-fortunately, coronagraphic observations, which provide CMEs speeds, are subject to the projection effects. In this paper, we presented a new way to estimate the radial speeds of CMEs. For this purpose we introduced the asymmetry ratio (ASR), the parameter obtained from the maximum and minimum speeds measured at different position angles for a given event. This allowed us to estimate realistic radial speeds. In Figure 2 and 3 we compared the travel time (TT) prediction using the projected and improved speeds. Respective correlation coefficients clearly show that the improved speed allows us to predict the TT with better accuracy. The standard error is about 2 hours smaller for the improved speeds than for the projected speeds. A similar situation appears when we compare correlation between speeds and source locations of CMEs and the strength of magnetic storms ($D_{ST}$ index). The correlation coefficients are about two times bigger for improved speeds in comparison with the projected speeds (Figure 4 and 5). It is important to compare our results to previous considerations. Xie *et al.* (2006) calculated absolute differences between predicted (using the ESA model, Gopalswamy *et al.* 2005b) and observed shock travel times for the cone models (XOL, MGY, ZPL). They found that the mean errors for those models were: 6.5, 12.8 and 9.2 hours, respectively. Recently, Michalek *et al.* (2007) calculated the same, but for the asymmetric cone model (error=8.4 hours). In the present considerations, the mean difference between predicted (using polynomial fit from Figure 2) and observed shock travel times is 10 hours, similar to the error from complicated cone models. Many authors demonstrated that the initial speeds of CMEs are correlated with the $D_{ST}$ index but the correlation coefficient were not significant because they applied the plane of the sky speeds. Using the MGY model, Michalek *et al.* (2006) showed that the correlation between the space speed of HCMEs and $D_{ST}$ index could be much more significant (correlation coefficient was ∼0.60). Recently, Michalek, Gopalswamy and Yashiro (2007) showed that for the asymmetric cone model the correlation coefficient for the wester events could be very significant (0.85). In the present study we found that correlation coefficient is 0.65 for the western events. Because in the present study we performed more accurate measurements of the projected speeds we got better results in comparison to the (MGY) model. Unfortunately, we do not involve the width of CMEs so we can only estimate the radial speeds of CMEs and our results are not as accurate as those of the asymmetric cone model (M). There are two important advantages of this method. Firstly, this technique is very fast. To get the ASR we need only to determine the height-time plots from coronagraphic observations. We used the average maximum and minimum speeds obtained from four successive measurements running around the occulting disk. Secondly, we do not apply any model or assumptions. The method also has some weak points. Faint HCMEs can not be used for the study, because it is difficult to get the height-time plots around the occulting disk. Fortunately poor events are generally not geoeffective so they are not of immediate concern. In contrast to the cone models we can not estimate the widths of HCMEs. Neglecting this parameter may lead to overestimation of the radial speeds. Although we obtained very accurate results using very simple considerations it is important to note that CMEs have very complicated 3D structure (Cremades and Bothmer, 2004) and more factors may have to be included when determining their geo-effectiveness.

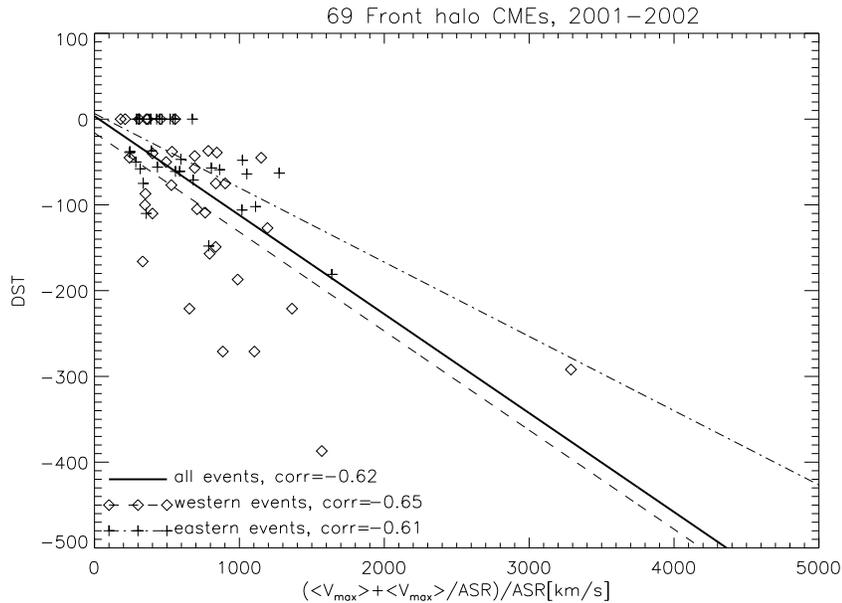

*Figure 5.* The scatter plot of $(< V_{max} + < V_{max} > /ASR)/ASR$ versus $D_{ST}$ index. Diamond symbols represent events originating from the western hemisphere and cross symbols represent events originating from the eastern hemisphere. The solid line is a linear fit to all the data points, the dot-dashed line is a linear fit to the eastern events, and the dashed line is a linear fit the western events.

## Acknowledgements

Work supported by NASA LWS T&T and SR&T grants. Work done by GM was supported in part by *MNiSW* through the grant N203 023 31/3055. SY was supported in part by NASA (NNG05GR03G).

Table .
**List of halo CMEs with the determined asymmetry ratio ASR**

| DATA | $V_{max}$ | $V_{min}$ | ASR | V | Location | $D_{ST}$ | TT |
|---|---|---|---|---|---|---|---|
| | km/s | km/s | | km/s | | nT | hours |
| 2001/01/10 | 809 | 359 | 2.24 | 832 | N13E36 | - | - |
| 2001/01/20a | 850 | 384 | 1.98 | 839 | S07E40 | -61 | 66 |
| 2001/01/20b | 1510 | 451 | 3.34 | 1507 | S07E46 | -61 | 64 |
| 2001/01/28 | 1090 | 311 | 3.50 | 916 | S04W59 | -40 | 68 |
| 2001/02/10 | 830 | 224 | 3.71 | 956 | N37W03 | -50 | 69 |
| 2001/02/11 | 1084 | 368 | 2.94 | 1183 | N24W57 | -50 | 50 |
| 2001/03/19 | 557 | 315 | 1.77 | 389 | S20W00 | -75 | 81 |
| 2001/03/24 | 973 | 325 | 2.99 | 906 | N16E22 | -56 | 72 |
| 2001/03/25 | 1020 | 276 | 3.69 | 677 | N16E25 | -87 | 67 |
| 2001/03/28 | 598 | 397 | 1.43 | 519 | N18E02 | - | - |
| 2001/03/29 | 1021 | 773 | 1.32 | 942 | N20W19 | -387 | 38 |
| 2001/04/01 | 1462 | 262 | 5.58 | 1475 | S22E90 | - | - |
| 2001/04/05 | 1447 | 607 | 2.38 | 1390 | S24E50 | -59 | 49 |
| 2001/04/06 | 1320 | 791 | 1.66 | 1270 | S21E31 | -63 | 64 |
| 2001/04/09 | 1118 | 581 | 1.92 | 1192 | S21W04 | -271 | 46 |
| 2001/04/10 | 2175 | 805 | 2.70 | 2411 | S23W09 | -271 | 33 |
| 2001/04/11 | 1179 | 396 | 2.97 | 1103 | S22W27 | -77 | 42 |
| 2001/04/12 | 1389 | 622 | 2.23 | 1184 | S19W43 | -75 | 35 |
| 2001/04/26 | 961 | 416 | 2.31 | 1006 | N20W05 | -47 | 41 |
| 2001/08/14 | 649 | 426 | 1.52 | 618 | N16W36 | -105 | 67 |
| 2001/08/25 | 1488 | 422 | 3.52 | 1433 | S17E34 | - | - |
| 2001/09/11 | 755 | 305 | 2.47 | 791 | N13E35 | - | - |
| 2001/09/24 | 2439 | 829 | 2.94 | 2402 | S16E23 | -102 | 34 |
| 2001/09/28 | 844 | 496 | 1.70 | 846 | N10E18 | -148 | 59 |
| 2001/10/01 | 1351 | 276 | 4.89 | 1405 | S24W81 | -166 | 55 |
| 2001/10/09 | 1030 | 467 | 2.20 | 973 | S28E08 | -71 | 54 |
| 2001/10/19a | 906 | 593 | 1.52 | 901 | N15W26 | -187 | 48 |
| 2001/10/19b | 605 | 166 | 3.64 | 558 | N16W18 | - | - |
| 2001/10/22 | 1315 | 563 | 2.33 | 1336 | S21E18 | -57 | 35 |
| 2001/10/25 | 1080 | 530 | 2.03 | 1090 | S16W21 | -157 | 58 |
| 2001/11/01 | 457 | 138 | 3.30 | 453 | N12W23 | - | - |
| 2001/11/03 | 555 | 247 | 2.24 | 457 | N04W20 | - | - |
| 2001/11/04 | 1920 | 1729 | 1.11 | 1810 | N06W18 | -292 | 33 |
| 2001/11/17 | 1553 | 702 | 2.21 | 1379 | S13E42 | -48 | 54 |
| 2001/11/21 | 543 | 295 | 1.84 | 518 | S14W19 | - | - |
| 2001/11/22a | 1474 | 490 | 3.00 | 1443 | S25W67 | -221 | 35 |
| 2001/11/22b | 1649 | 872 | 1.86 | 1437 | S19W42 | -221 | 33 |
| 2001/11/28 | 679 | 361 | 1.87 | 500 | N04E16 | - | - |
| 2001/12/13 | 853 | 520 | 1.63 | 864 | N16E09 | -39 | 87 |
| 2001/12/14 | 1402 | 211 | 6.64 | 1506 | N07E86 | -39 | 39 |
| 2001/12/28 | 1920 | 362 | 5.31 | 2216 | S24E90 | - | - |
| 2002/01/04 | 1205 | 357 | 3.36 | 896 | N30E75 | - | - |
| 2002/01/14 | 1510 | 354 | 4.94 | 1492 | S28W90 | - | - |
| 2002/02/20 | 835 | 236 | 3.50 | 952 | N12W72 | - | - |
| 2002/03/10 | 1410 | 245 | 5.75 | 1429 | S22E90 | - | - |
| 2002/03/11 | 953 | 234 | 4.07 | 950 | S15E45 | - | - |
| 2002/03/14 | 961 | 299 | 3.21 | 961 | S23E57 | -37 | 93 |
| 2002/03/15 | 978 | 514 | 1.91 | 957 | S08W03 | -37 | 62 |
| 2002/03/18 | 958 | 393 | 2.43 | 989 | S10W20 | - | - |
| 2002/03/22 | 1754 | 297 | 5.98 | 1750 | S10W90 | -100 | 39 |
| 2002/04/15 | 697 | 624 | 1.11 | 720 | S15W01 | -127 | 56 |
| 2002/04/17 | 1242 | 571 | 2.17 | 1240 | S14W34 | -149 | 60 |
| 2002/04/21 | 2374 | 560 | 4.24 | 2393 | S14W84 | -57 | 49 |
| 2002/05/07 | 637 | 255 | 2.50 | 720 | S10E25 | -110 | 79 |
| 2002/05/08 | 523 | 265 | 1.97 | 614 | S12W07 | -110 | 69 |
| 2002/05/16 | 612 | 228 | 2.67 | 600 | S23E15 | -58 | 67 |
| 2002/05/22 | 1612 | 565 | 2.85 | 1557 | S30W34 | -109 | 32 |
| 2002/07/15 | 1043 | 337 | 3.08 | 1151 | N19W01 | - | - |
| 2002/07/18 | 926 | 389 | 2.38 | 1099 | N19W30 | -38 | 94 |
| 2002/07/20 | 1640 | 216 | 7.59 | 1941 | S13E90 | -38 | 32 |
| 2002/07/23 | 1717 | 327 | 5.25 | 2285 | S13E72 | - | - |
| 2002/07/26 | 756 | 236 | 3.20 | 818 | S19E26 | - | - |
| 2002/08/16 | 1466 | 691 | 2.12 | 1585 | S14E20 | -106 | 54 |
| 2002/08/22 | 928 | 201 | 4.58 | 998 | S07W62 | -45 | 106 |
| 2002/08/24 | 1886 | 804 | 2.34 | 1913 | S02W81 | -45 | 57 |
| 2002/09/05 | 1739 | 1028 | 1.69 | 1748 | N09E28 | -181 | 45 |
| 2002/11/09 | 1661 | 524 | 3.16 | 1838 | S12W29 | -43 | 48 |
| 2002/11/24 | 964 | 634 | 1.52 | 1077 | N20E35 | -64 | 50 |
| 2002/12/19 | 1050 | 549 | 1.91 | 1092 | N15W09 | -75 | 68 |